\documentstyle[12pt]{article}

\def\tr{\mathop{\rm tr}}
\def\diag{\mathop{\rm diag}\nolimits}
\def\res{\mathop{\rm res}}
\def\U{{\cal U}}
\def\X{\hat m_q}
\def\bbar#1{\,\overline{\!#1\!}\,}

\let\dsize=\displaystyle

\let\t=\dagger

\begin{document}
\begin{titlepage}
\begin{center} 
\vspace*{1in}
{\LARGE {\bf Extended chiral group and scalar diquarks}} 
\vskip .5in
{\bf Yuri Novozhilov$^{1)}$, Andrei Pronko$^{2)}$,
Dmitri Vassilevich$^{3)}$ \\ and Alexander Korotkov$^{1)}$}
\vskip .5in
\begin{flushleft}
{\footnotesize {\it \hspace{2cm} 
1) Department of Theoretical Physics, 
St.~Petersburg State University\\ \hspace{2cm}
2) St.~Petersburg Department of Steklov Mathematical Institute\\ \hspace{2cm}
3) Physics Department, University of Leipzig}}
\end{flushleft}
\end{center}
\vskip 1in
\begin{abstract}
\normalsize

We introduce extended chiral  transformation, which depends both
on pseudoscalar and diquark fields as parameters and determine its group
structure. Assuming soft symmetry breaking in diquark sector, bosonisation
of a quasi-Goldstone $ud$-diquark is performed. In the chiral limit the
$ud$-diquark mass is defined by the gluon condensate,
$m_{ud}\approx 300 MeV$. The diquark charge radius is
$\langle r^2_{ud}\rangle^{1/2}\approx 0.5 fm$.
We consider also the flavour triplet of scalar diquarks
$(ud)$, $(us)$ and $(ds)$ together with pseudoscalar mesons
and calculate diquark masses and decay constants in terms of
meson parameters and the gluon condensate.
\end{abstract}
\end{titlepage}

\section{Introduction.}

Diquarks, which were introduced almost three decades ago [1],
became now efficient tool for studying various processes in hadron physics
(see e.g. [2,3] and reviews [4]). The diquark model was analysed from
various points of view. However, the complete picture is still lacking.

It was suggested by Dosch et al [3] that wave function of pion and
$ud$-diquark  are the same at the origin. It was also shown [3] that the
diquark decay constant following from this suggestion is very close to that
estimated from the QCD sum rules.

In this paper we propose to go a bit further: we suppose that the
similarity of wave functions of pion and diquark  is due to common origin
as parameters of a certain anomalous transformation which does not
preserve the measure of the quark path integral. While at the classical
level the chiral symmetry is broken by quark mass, the extended chiral
(E$\chi$) symmetry is broken by quark mass and gluon fields. E$\chi$-group
is $U(2N)$ for $N$ internal degrees of freedom, $N=N_c N_f$. Non-anomalous
(measure preserving) generators span the Lie algebra of $O(2N)$, anomalous
generators belong to the coset $U(2N)/O(2N)$. Anomalous generators describe
chiral rotations and transformations with diquark variables (``diquark''
rotations), non-anomalous part consists out of gauge transformations and
combined chiral ``diquark'' rotations [5].

We assume that E$\chi$-symmetry breaking due to quark masses and gluon
fields is soft in the sense that the action for bosonised diquark fields
can be obtained by integrating corresponding E$\chi$-anomaly. Colorless
chiral fields after bosonisation give rise to Goldstone particles --
pseudoscalar mesons. We suggest that at low energies bosonised diquark
parameters of E$\chi$-transformations with quantum numbers of lightest
$J^P=0^+$ $ud$-diquark can be treated as a Goldstone-like particle.
Therefore, in bosonisation we restrict ourselves  to the case of
E$\chi$-transformations
with $ud$-diquark fields. The E$\chi$-group in this case is
$SU(4)$, non-anomalous transformations are just gauge transformations
$SU(3)\times U(1)$ and the diquark Goldstone degrees of freedom belongs to
$CP^3=SU(4)/SU(3)\times U(1)$ [5].

Our analysis shows that the $ud$-diquark introduced
{\it a la} Goldstone becomes massless in the limit of vanishing
gluon condensate and current quark masses. Furthermore, we calculate the
diquark mass and charge radius for the actual  value of gluon condensate.
The obtained values fall into the region allowed  in other models [4].
Note, that our approach is a direct generalization of chiral bosonisation
scheme [6] for the case of new anomalous transformation. We introduce no
new parameters.

In the section four we investigate $E\chi$ - action for $N_f=3$
taking into account quark masses [7]. We obtain expressions which
determine masses and decay constants of scalar diquarks
$(ud)$, $(us)$ and $(ds)$. We also derive relations connecting
masses and decay constants of scalar diquarks and pseudoscalar
mesons $\pi$ and $K$. By these calculations we demonstrate
advantages of $E\chi$ group. Though diquarks are not physical
particles, their masses and decay constants should be considered at
the same level as quark masses. Applications to nucleon and 
scattering will be considered in a separate paper.

\section{Group  structure of E$\chi$-transformations.}

It was demonstrated [8] that in order to consider
quark-antiquark and quark-quark composites on equal footings one should
introduce eight-component spinors $\Psi$ constructed from ordinary Dirac
spinors $\psi$
\begin{equation} \label{Psi}
\Psi=\left( \begin{array}{l} \psi \\
									 \bar\psi^T
\end{array} \right)
\end{equation}
The quark lagrangian can be rewritten in the form
\begin{equation} \label{lag}
{\cal L} = \frac{1}{2} \Psi^T \hat F \Psi,\qquad
\hat F=\left(
\begin{array}{cc}  C\Phi & -D^T \\
											  D     &  \overline{\Phi}C
\end{array} \right) \qquad
F=-F^T
\end{equation}
where $D$ is the Dirac operator
$D=i\gamma^\mu(\partial_\mu+v_\mu+\gamma_5 a_\mu)$, ``${}^T$'' means
transposition and $\Phi=\gamma^\mu(\phi_{5\mu}+\gamma_5\phi_\mu)$,
$\overline\Phi=\gamma_0\Phi^+\gamma_0$. We have introduced various
external fields $v_\mu,a_\mu,\phi_\mu$ and $\phi_{5\mu}$ generating both
$\bar\psi \psi$ and $\psi\psi$ composites. $C$ is charge conjugation
matrix. The quark path integral becomes
\begin{eqnarray} \label{Z-psi}
Z_\psi & = & \int {\cal D}\Psi \exp i\int d^4x {\cal L} = (\det\hat G)^{1/2}
\nonumber\\
\hat G &=& \left(\begin{array}{cc} D & \overline{\Phi}\\
																		  \Phi   & D_c
\end{array}\right), \qquad
D_c=C^{-1}D^T C,
\nonumber\\
\hat F &=& \left(\begin{array}{cc}
0 & -1\\
1 &  0
\end{array}\right)
\left(\begin{array}{cc}
1 & 0 \\
0 & C^{-1}
\end{array}\right)
\hat G
\left(\begin{array}{cc}
1 & 0\\
0 & C
\end{array}\right)
\end{eqnarray}
where the operator $\hat G$ is $\gamma_0$-hermitian.
A similar construction was considered by Ball [9] for Majorana spinors.

The lagrangian (\ref{lag}) is invariant under the following
transformations
\begin{equation} \label{Omega}
\delta\Psi=-\Omega\Psi, \qquad
\Omega=
\left(\begin{array}{cc}
\alpha+\gamma_5\chi & (\xi+\gamma_5\omega)C\\
(-\xi^*+\gamma_5\omega^*)C & \alpha^*-\gamma_5\chi^*
\end{array}\right)
\end{equation}
provided  external fields in the operators $\hat G$ and $\hat F$
transform according to the rules
\begin{eqnarray} \label{Xi-Theta}
\hat F \rightarrow  \hat F' & = & \exp\Omega^T \, \hat F \, \exp\Omega
\nonumber\\
\hat G  \rightarrow \hat G' & = & \exp(-\Xi+\gamma_5\Theta)\, \hat G  \,
																									\exp(\Xi+\gamma_5\Theta)
\nonumber\\
\Xi & = & \left(\begin{array}{rr}
\alpha   &  \xi  \\
\xi^*   &  \alpha^*
\end{array}\right)
\qquad
\Theta=\left(\begin{array}{rr}
\chi      & \omega \\
-\omega^* & -\chi^*
\end{array}\right).
\end{eqnarray}
The matrices  $\alpha$ and $\chi$ are antihermitian, $\xi$ is antisymmetric
and $\omega$ is symmetric in internal indices. The transformations
(\ref{Omega}) do not destroy the structure (\ref{Psi}) of the
eight-component spinor $\Psi$. These transformations can be absorbed in
transformations of background fields.

Due to the noninvariance of the measure, only part of the transformations
(\ref{Omega}) do not change the path integral (\ref{Z-psi}).
These are the transformations generated by $\Xi$. The generators $\Theta$
lead to quantum anomalies. The operators $\alpha$ generate gauge
transformations, $\chi$ describe chiral rotations, the anomalous
transformations $\omega$ include fields with  diquark quantum numbers,
the generators $\xi$ are needed for closure of the algebra.

The matrix commutator
\begin{equation} \label{com}
[\Xi(\alpha_1,\chi_1)+\gamma_5 \Theta(\xi_1,\omega_1),
\Xi(\alpha_2,\chi_2)+\gamma_5 \Theta(\xi_2,\omega_2)]
=\Xi(\alpha_3,\chi_3)+\gamma_5 \Theta(\xi_3,\omega_3)
\end{equation}
induces the following Lie structure
\begin{eqnarray} \label{laws}
\alpha_3&=&[\alpha_1,\alpha_2]+[\chi_1,\chi_2]+\xi_1\xi_2^*
-\xi_2\xi_1^*-\omega_1\omega_2^*+\omega_2\omega_1^*,
\nonumber\\
\chi_3&=&[\alpha_1,\chi_2]-[\alpha_2,\chi_1]-\xi_1\omega_2^*
-\omega_2\xi_1^*+\xi_2\omega_1^*+\omega_1\xi_2^*,
\nonumber\\
\xi_3&=&\alpha_1\xi_2+\xi_1\alpha_2^*-\alpha_2\xi_1-\xi_2\alpha_1^*
+\chi_1\omega_2-\omega_1\chi_2^*-\chi_2\omega_1+\omega_2\chi_1^*,
\nonumber\\
\omega_3&=&\alpha_1\omega_2+\chi_1\xi_2-\xi_1\chi_2^*+\omega_1\alpha_2^*
-\alpha_2\omega_1-\chi_2\xi_1+\xi_2\omega_1^*-\omega_2\xi_1^*.
\end{eqnarray}
One can verify that the composition laws (\ref{laws}) are induced also
by the
matrix commutator without $\gamma_5$
\begin{equation} \label{out-5}
[\Xi(\alpha_1,\chi_1)+\Theta(\xi_1,\omega_1),
\Xi(\alpha_2,\chi_2)+ \Theta(\xi_2,\omega_2)]
=\Xi(\alpha_3,\chi_3)+\Theta(\xi_3,\omega_3)
\end{equation}
This means that the Lie algebras (\ref{com}) and (\ref{out-5}) are
isomorphic. For the case of $N$ internal degrees of freedom, $N=N_c N_f$,
and maximally extended algebra (i.e. when $\alpha, \chi, \xi$ and $\omega$
are all matrixes satisfying the above mentioned hermiticity and symmetry
properties), the $\Xi+\Omega$ form the space of hermitian matrices
$2N\times 2N$. Hence the algebra (\ref{out-5}) is $U(2N)$. The generators
$\Xi$ preserve symmetric non-degenerate bilinear form $O$
\begin{equation} 
O=\left(
\begin{array}{cc}
0 & 1 \\
1 & 0
\end{array}
\right), \qquad \Xi O + O \Xi^T=0.
\end{equation}
Consequently the non-anomalous generators $\Xi$ span the Lie algebra of
$O(2N)$ and the anomalous generators  belong to the coset $U(2N)/O(2N)$.
The generators $\alpha $ and $\omega$ preserve symplectic form
\begin{equation} 
  \Sigma=\left(
\begin{array}{cc}
0 & 1 \\
-1 & 0
\end{array}
\right)
\end{equation}
and  thus span the Lie algebra of the subgroup $Sp(N)$. The generators
obviously form $U(N)$.

In principle, any transformation of $\Theta$ could be related to a Goldstone
particle, whose dynamics is governed by quantum anomalies. However in
realistic models most of the symmetries (\ref{Omega}) are broken already at
classical level by the presence of  quark masses and gluon fields, and
the vector field prescribed by the transformation rules (\ref{Xi-Theta}).
Only colorless chiral fields $\chi$ are definitely interpreted as
pseudoscalar mesons. Some other states were also considered in literature
[8]. We suggest, that at certain energy scale the fields $\omega$ with
quantum numbers of lightest $J^P=0^+$ $ud$-diquarks can also generate
Goldstone-like particles. In what follows we shall restrict ourselves to the
transformations
\begin{equation} \label{omega}
\omega=(1/f_\omega)\omega_c(i\sigma_2)_{jk}\epsilon_{abc}
\end{equation}
corresponding to $0^+$ $ud$-diquarks where j,k are flavor  and a,b,c are
color indices.

In this special case transformations close in a smaller group. To see this
one should exclude the $i\sigma_2$ in the same way, as it was done
previously with $\gamma_5$, and use the commutation relations
(\ref{laws}). One can obtain that after removing $i\sigma_2$ and $\gamma_5$
the algebra becomes  formally equivalent to that generated by the $\alpha$
and $\xi$ operators in the case $N=3$. Hence, the complete group is
$O(6)\sim SU(4)$, and the non-anomalous transformations, that are now
represented by $\alpha$ generators, belong to $U(3)\sim SU(3)\times U(1)$.
The anomalous (Goldstone) diquark degrees of freedom belong to the complex
projective space $CP^3=SU(4)/SU(3)\times U(1)$. The same result could be
obtained in a straightforward but tedious way by computing matrix
commutators in an appropriate basis.

As a consistency check we shall demonstrate
that the diquark mass vanishes for zero gluon condensate and zero current
quark masses. We shall also compute the diquark mass and charge radius for
actual value of gluon condensate.

\section{The diquark bosonisation. Gluon condensate as a
source of diquark mass.}

To define the diquark  parameters we should regularize the quark path
integral. We also need a method of extracting a non-invariant part of the
path integral corresponding to anomalous transformations.

To reduce possible regularization dependence [11] we shall use exactly the
same scheme [6] which was developed for chiral bosonisation and generalized
[8] for the presence of diquark variables. Since the parameters of this
scheme were defined through chiral dynamics, we will be able to compare our
results for diquark with pion physics directly.

The basic object is the quark path integral over low scale region
\begin{eqnarray} \label{Z^L}
Z_{\psi}^L &=& (\det\{\hat G \theta(1-(\hat G-M)^2/\Lambda^2)) \})^{1/2}
\nonumber\\
\theta(x) &=& \int_{-\infty}^{\infty} dt\frac{\exp(ixt)}{2\pi i(t-i0)}.
\end{eqnarray}
The parameters $\Lambda$ and $M$ are defined below. The functional
(\ref{Z^L}) can be represented in the form
\begin{eqnarray} 
Z_\psi^L &=& (Z_\psi^L Z_{inv}^{-1}) Z_{inv}
\nonumber\\
Z_{inv}^{-1} &=& \int {\cal D} \Theta (Z_\psi^L(\Theta))^{-1}
\end{eqnarray}
where we integrate over anomalous transformations $\Theta$,
${\cal D} \Theta$ is invariant measure on the corresponding coset space and
$(Z_\psi^L(\Theta))$ is the path integral (\ref{Z^L}) with background
fields transformed as in eq.(\ref{Xi-Theta}). The $Z_{inv}$ does not depend
on degrees of freedom described by $\Theta$. Hence all information over
$\Theta$-noninvariant processes is contained in $(Z_\psi^L Z_{inv}^{-1})$
and the effective action for $\Theta$ can be defined as
\begin{equation} 
(Z_\psi^L Z_{inv}^{-1})=\int {\cal D} \Theta  \exp(iW_{eff}(\Theta))
\end{equation}
The effective action is obtained by integration of the corresponding
anomaly ${\cal A}(x)$
\begin{eqnarray} \label{Weff}
{\cal A}(x;\Theta) &=&
\frac{1}{i}\frac{\delta\ln Z_\psi^L(\Theta)}{\delta\Theta}
\nonumber\\
W_{eff}(\Theta) &=& -\int d^4x \int_{0}^{1} ds {\cal A}(x;s\Theta)\Theta(x)
\end{eqnarray}
Previously [5] this method was applied
to $\pi$-mesons.

An important dynamical information is contained in parameters
$\Lambda$ and $M$. 
These parameters define a low-energy region of the model [6].  
The quark and the gluon condensates 
are related to $\Lambda$ and $M$ in the following way 
\begin{equation} 
C_g=\frac{3N_c}{2\pi^2}(6\Lambda^2M^2-\Lambda^4-M^4),\qquad
C_q=-\frac{N_c}{2\pi^2}(\Lambda^2M-\frac{1}{3}M^3).
\end{equation} 
It was found also that these parameters are
related to the pion decay constant by the formula
\begin{equation} 
f_\pi^2=\frac{N_c}{2\pi^2} (\Lambda^2-M^2)
\end{equation}
We will not report here details of computations
 of $W_{eff}(\omega)$ for
the case $\Theta=\omega$, where $\omega$ is given by (\ref{omega}).
They can be performed in same manner as
in the papers [5,6]. Neglecting all external fields except vector gauge
fields
\begin{equation} 
v_\mu=-iQA_\mu+\frac{\lambda^a}{2i}G_\mu^a,
\qquad
Q=
\left(\begin{array}{cc}
2/3 & 0\\
0   & -1/3
\end{array}\right)
\end{equation}
where $A_\mu$ is electromagnetic field, $G_\mu^a$ are gluons
and taking zero current quark masses we
obtain in quadratic order of $\omega$
\begin{eqnarray} \label{W_eff}
\lefteqn{
W_{eff}(\omega)=
\frac{1}{192\pi^2{f_\omega}^2} \tr_{(c,f)} \Biggl\{
6(\Lambda^2-M^2)[D_\mu,\omega^*][D^\mu,\omega]
}
\nonumber\\&&
+[D_\mu,[D^\mu,\omega^*]] [D_\nu,[D^\nu\omega]]
+2[D_\mu,F^{\mu\nu}]
(\omega[D_\nu,\omega^*]+[D_\nu,\omega]\omega^*)
\nonumber\\&&
+(F^{\mu\nu}F_{\mu\nu}\omega\omega^*
-F^{\mu\nu}\omega F_{\mu\nu}^T\omega^*)
\Biggr\}
\end{eqnarray}
where
$[D_\mu,\omega] = (\partial_\mu\omega)+v_\mu\omega+\omega v_\mu^T,
[D_\mu,\omega^*] = (\partial_\mu\omega^*)-v_\mu^T\omega^*-\omega^* v_\mu $
and
$F_{\mu\nu} = (\partial_\mu v_\nu)-(\partial_\nu v_\mu)+[v_\mu,v_\nu]$.
>From (\ref{W_eff}) we see that the mass of $ud$-diquark
$\omega$ is defined
by the gluon condensate $\langle G_{\mu\nu}^2\rangle$ ($N_c=3$)
\begin{equation} \label{M-ud}
M_\omega^2=-2\pi^2 f_\pi^2
+\sqrt{4\pi^4 f_\pi^4+\frac{\langle G_{\mu\nu}^2\rangle}{12}}
\end{equation}
and vanishes when $\langle G_{\mu\nu}^2\rangle \rightarrow 0$.
For derivation of (\ref{M-ud})
we used
\begin{equation}   
\langle G_{\mu\nu}^a G^{b \mu\nu}\rangle=
\frac{1}{8}\delta^{ab}\langle G_{\mu\nu}^2 \rangle
\end{equation}
For $\langle G_{\mu\nu}^2\rangle = (365 MeV)^4$ we get $M_\omega\approx
300 MeV$. The correction of this evaluation due to quark masses
is provided by $M_\omega^2(m_q\ne 0)=M_\omega^2(m_q=0)+m_\pi^2$.
This gives $M_\omega^2(m_q\ne 0)\approx 340 MeV$, which falls into the
region allowed in the other models [4], though lies close to the lower
boundary. $f_\omega$ is defined by requirement that the residue of the
diquark propagator at $k^2=M_\omega^2$ is unity,
\begin{equation} 
f_\omega^2=\frac{\Lambda^2-M^2}{2\pi^2}+ \frac{1}{6\pi^2}M_\omega^2
\end{equation}

The coefficient before the term
$\partial^2A^\mu
(\omega^*(\partial_\mu\omega)-(\partial_\mu\omega^*)\omega)$
allows us to evaluate the mean
square radius of diquark charge distribution
\begin{equation} 
\langle r^2 \rangle^{1/2} \approx  0.5 fm
\end{equation}
This value is also compatible with other data [4] for diquark effective
radius.

Our desire to describe diquarks as  a quasi-particle similar to
$\pi$ meson has more
than aesthetic grounds. This model allows to explain relatively low mass of
the scalar diquark and include diquark variables in framework of current
algebra and chiral perturbative theory.
As far as we were able to
verify, this suggestion  does not lead to any contradictions. We
obtained quite sensible results for diquark mass and charge radius.
The model has no free parameters. All this indicates that
broken E$\chi$-symmetry deserves further investigations.

\section{E$\chi$ -- action for $SU(3)$ flavour structure.}

It was demonstrated in the section one that in order to consider
quark-antiquark and quark-quark composites on an equal footing
one should introduce an eight-component spinor $\Psi$ constructed
from ordinary Dirak spinors $\psi$
$$
\Psi =\left(
\begin{array}{c}
\psi \\ \bar{\psi}^T
\end{array}
\right)
$$
In this section we consider $SU(3)$ flavour structure
of the fermion field and transformations of the form:
\begin{equation}
\delta\Psi=\Theta\Gamma_5\Psi,\ \ \ \ \ \
\Theta=
\left(
\begin{array}{ll}
\chi & C\omega \\ C\omega^* & \chi^T 
\end{array}
\right),
\end{equation}
where $\Gamma_5$ is a block-diagonal matrix constructed
from the $\gamma_5$ matrixes,
$\chi$ contains pseudoscalar meson states ($J^P=0^-$) 
and $\omega$ contains scalar diquark states ($J^P=0^+$) 
\begin{equation} \label{1836}
\chi=\chi_{(1)},\qquad
\omega=\omega_{(\bar 3)}.
\end{equation}
The representations of the colour group $SU(3)_á$ are
indicated explicitly. Thus,
$\chi_{(1)}$ contains only flavour octet 
(the singlet component connected with $U_A(1)$--current is
excluded as it is in the chiral theory), 
$\omega_{(\bar 3)}$ is antitriplet ($\bar 3_c\otimes \bar 3_f$). 

Let us introduce the $E\chi$-field according to (\ref{Xi-Theta})
\begin{equation}\label{Echi}
\U=-\exp i\Theta
\end{equation}

Introducing scalar diquarks on the equal footing with
pseudoscalar mesons in the field $\U$ can
be considered as an alternative for the method, based
on composite quark-quark fields  [12].
In the present approach states with quantum numbers $J^P=0^+$
in quark-quark sector occupy the same place as states with $J^P=0^-$
in quark-antiquark sector (the space part of the wave function
is symmetrical, fermions are in $S$-wave).
The distinctions appear at dynamical level by 
analysis of the effective action. In particular, diquark
states are massive in the chiral limit.

Let us present E$\chi$--action (\ref{Weff})
It would be convenient to write
$W_{{\rm eff}}=W+W'$ where $W$ is $SU(N_f)_L\otimes SU(N_f)_R$
invariant part of E$\chi$--action (zero order of quark masses)
and part $W'$ comes from soft chiral 
$SU(N_f)_L\otimes SU(N_f)_R$ symmetry breaking 
(due to quark masses ).

The Lagrangian corresponding to the action $W$ has the standard form [6]
\begin{eqnarray} \label{L}
\lefteqn{
{\cal L}=
\frac{\Lambda^2-M^2}{32\pi^2}\tr_{(b,c,f)}(D_\mu \U)\, (D^\mu \U)^\t
+\frac{1}{192\pi^2}\tr_{(b,c,f)}\biggl\{(D_\mu^2 \U)\, (D_\nu^2 \U)^\t
}
\nonumber\\ &&
+\frac{1}{2}(D_\mu \U)\, (D_\nu \U)^\t\, 
(D^\mu \U)\, (D^\nu \U)^\t
-\left((D_\mu \U)\, (D^\mu \U)^\t \right)^2
\nonumber\\ &&
+2(D_\mu F^{\mu\nu})\, \left((D_\nu \U)\, 
\U^\t +(D_\nu \U)^\t \, \U \right)
-\frac{1}{2}[F_{\mu\nu}, \U] [F^{\mu\nu}, \U^\t ]\biggr\}
\nonumber\\ ,
\end{eqnarray}
The Wess - Zumino -  Witten action will not be considered here
The action  $W'$ containing 
terms with quark masses is determined by Lagrangian
\begin{eqnarray} \label{L'}
\lefteqn{
{\cal L}'=
-{1\over 8\pi^2}
\tr_{(b,c,f)}
\Biggl\{\left(\Lambda^2 M - {M^3 \over 3}\right)\, \X ({\cal U}+{\cal U}^\t)
}
\nonumber\\ &&
+{\Lambda^2-M^2 \over 4}({\U} \X {\U} 
\X + \X {\U}^\t \X {\U}^\t)
-{M\over 2} \X(D_\mu D^\mu {\U} + D_\mu D^\mu {\U}^\t)
\Biggr\}
\nonumber\\ &&
+ O(\X^3),
\end{eqnarray}
where matrix $\X$ describes quark masses
\begin{equation} 
\X=1_b\otimes 1_c\otimes\diag_f(m_u, m_d, m_s).
\end{equation}
The covariant derivative $D_\mu$ is
$(D_\mu *)=(\partial_\mu *)+[V_\mu, *]$,  $V_\mu$
and its field strength $F_{\mu\nu}$ 
are block - diagonal matrixes depending only on gluon field $G_\mu$
and its field strength correspondinly
\begin{equation} 
V_\mu= \pmatrix{ G_\mu  & 0 \cr 0  & -G_\mu^T},\qquad
F_{\mu\nu} =\pmatrix{G_{\mu\nu}& 0 \cr 0  & -G_{\mu\nu}^T}.
\end{equation}

Let us note that the global flavour invariance $SU(3)_L\otimes SU(3)_R$
of lagrangian (\ref{L}) is realized as follows
\begin{equation} 
\U\to
\pmatrix{\ell &0\cr 0& r^{-1\,T}} \U \pmatrix{r^{-1} &0 \cr 0& \ell^T},
\qquad
\ell\in SU(3)_L,\quad r\in SU(3)_R.
\end{equation}
It is easy to see that though field $\U$ has a complicated structure
(when $N_f=3$ the field  $\U$ is an element of
$SU(18)/O(18)$ coset space (see section two)),
global symmetries of lagrangian ${\cal L}$ wider than 
$SU(3)_L\otimes SU(3)_R$ are not possible in the presence of
coloured gluon fields in the covariant derivative.

Let us calculate masses and decay constants
of $\bar 3$ scalar diquarks.
Only these fields can form colourless baryon states
with a third quark. For analysis of diquark
parameters we consider decay constants and masses
of pseudoscalar mesons
as parameters known from an experiment and
express dynamical characteristics of scalar diquarks
in terms of the gluon condensate and these parameters.

Let us suppose for simplicity that $m_u=m_d=:m$, i.e.
\begin{equation} \label{mms}
\hat m_q =1_b\otimes 1_c\otimes \diag(m,m,m_s)
\end{equation}
in (\ref{L'}). The matrix of pseudoscalar mesons is given by
\begin{equation} \label{pK}
\chi_{1_c} =\left(
\begin{array}{ccc}
\dsize\frac{\pi^0}{\sqrt{2}f_\pi} &
\dsize\frac{\pi^+}{f_\pi} &
\dsize\frac{K^+}{f_K} \\[10pt]
\dsize\frac{\pi^-}{f_\pi} &
-\dsize\frac{\pi^0}{\sqrt{2}f_\pi} &
\dsize\frac{K^0}{f_K} \\[10pt]
\dsize\frac{K^-}{f_K} & \dsize\frac{\bbar{K^0}}{f_K} & 0
\end{array}\right)
\end{equation}
where $f_\pi$ and
$f_K$ are decay constants of $\pi$-meson and $K$-meson
correspondingly, $f_\pi\approx$132Mev, $f_K\approx$165Mev.
The matrix $\omega_{(\bar 3)} $ in (\ref{1836}) containing fields of $\bar 3$
scalar diquarks has the following form:
\begin{equation} 
(\omega_{(\bar 3)})_{ij}^{ab}=
\varepsilon^{abc}\varepsilon_{ijk}\,\cdot\frac{\omega_k^c}{f_{\omega_k}},
\end{equation}
where $a,b,c$ are colour indexes, $i,j,k$ are flavour indexes and
$\omega_1^c$, $\omega_2^c$, $\omega_3^c$ are fields of $(ud)$, $(us)$,
$(ds)$ diquarks, $f_{\omega_1}$, $f_{\omega_2}$, $f_{\omega_3}$
are corresponding diquark decay constants.
Owing to residual $SU(2)_f$ invariance (\ref{mms})
$f_{\omega_1}=f_{\omega_2}\equiv f_{\omega_{1,2}}$.

Consider the propagators $S(p^2)$ (in the momentum representation).
The presence of the ``tachyonic'' term 
$(D^2\U)(D^2\U^\t)$ in (\ref{L}) leads to appearance of terms 
proportional to $p^4$.
Let us introduce a
real parameter $\alpha\in [0,1]$ replacing
$(D^2\U)(D^2\U^\t)$ by $\alpha(D^2\U)(D^2\U^\t)$
to facilitate discussing of the ``tachyonic'' term.
The solution of the equation
\begin{equation} \label{masseqn}
S^{-1}(p^2)=0
\end{equation}
gives us the mass.
The physical mass $m_{\rm phys}$ is 
the solution $p^2=m_{\rm phys}^2>0$. The other solution
$p^2=-m_{\rm tach}^2<0$ gives non-physical ``tachyonic'' pole of the propagator
$m_{\rm tach}\to\infty$ when $\alpha\to 0$.
The decay constant is determined by the condition
\begin{equation} \label{reseqn}
\res_{p^2=m_{\rm phys}^2} S(p^2)=1.
\end{equation}

We consider first the meson sector. The inverse propagator of
$\pi$-meson has the form
\begin{equation} \label{420}
S^{-1}_\pi (p^2)=\frac{\alpha}{4\pi^2 f_\pi^2}\, p^4
+\frac{A_\pi}{f_\pi^2}\, p^2 -\frac{B_\pi}{f_\pi^2},
\end{equation}
where $A_\pi$ and $B_\pi$ are constants of
dimensions two and four correspondingly. The propagator of  $K$-meson
has also the form (\ref{420}) with constants $f_K$, $A_K$, $B_K$ instead of
$f_\pi$, $A_\pi$, $B_\pi$. Masses and decay constants of $\pi$ and $K$
mesons can be expressed   
in the terms of constants $A_\pi$, $B_\pi$ and $A_K$, $B_K$ with the
help of (\ref{masseqn}) and (\ref{reseqn}).
Inversion of these expressions gives:
\begin{equation}
\begin{array}{l} 
\label{ABAB-1}
A_\pi
=\displaystyle{f_\pi^2 \left(1-\alpha\frac{m_\pi^2}{2\pi^2 f_\pi^2}\right)},\quad
B_\pi
=\displaystyle{m_\pi^2 f_\pi^2\left(1-\alpha\frac{m_\pi^2}{4\pi^2 f_\pi^2}\right)},\\
\\
A_K=\displaystyle{f_K^2 \left(1-\alpha\frac{m_K^2}{2\pi^2 f_K^2}\right)},\quad
B_K=\displaystyle{m_K^2 f_K^2\left(1-\alpha\frac{m_K^2}{4\pi^2 f_K^2}\right)}.
\end{array}
\end{equation}
On the other hand, it follows from expressions for
effective Lagrangians 
(\ref{L}) and (\ref{L'}) that
\begin{eqnarray} \label{ABAB-2}
\lefteqn{
A_\pi=\frac{3}{2\pi^2}\left(\Lambda^2-M^2+2Mm\right),
}
\nonumber\\
\lefteqn{
B_\pi=\frac{3}{2\pi^2}\biggl\{4\left(\Lambda^2 M-\frac{M^3}{3}\right) m
+4\left(\Lambda^2-M^2\right) m^2 + {\cal O}(m^3)\biggr\},
}
\nonumber\\
\lefteqn{
A_K=\frac{3}{2\pi^2}\left(\Lambda^2-M^2+M(m+m_s)\right),
}
\nonumber\\
\lefteqn{
B_K=\frac{3}{2\pi^2}\biggl\{
2\left(\Lambda^2 M-\frac{M^3}{3}\right)(m+m_s)
}
\nonumber\\ &&  \qquad\qquad\qquad
+\left(\Lambda^2-M^2\right) (m+m_s)^2
+ {\cal O}\left((m+m_s)^3\right)\biggr\}.
\end{eqnarray}
We remind that meson masses and decay constants
of mesons are supposed to be given parameters.
Thus, the relations (\ref{ABAB-1}) define the constants 
$A_\pi$, $B_\pi$, $A_K$, $B_K$. The expressions (\ref{ABAB-2})
are helpful when considering the diquark sector.

The inverse propagators of $(ds)$,
$(us)$ and $(ud)$ diquark fields $\omega_1^c$, $\omega_2^c$ and $\omega_3^c$
correspondingly can be presented in the form simular to (\ref{420}) 
\begin{equation}
\begin{array}{l} 
S^{-1}_{\omega_{1,2}} (p^2)
=\displaystyle{\frac{\alpha}{6\pi^2 f_{\omega_{1,2}}^2}\, p^4
+\frac{A_{\omega_{1,2}}}{f_{\omega_{1,2}}^2}\, p^2
-\frac{B_{\omega_{1,2}}}{f_{\omega_{1,2}}^2}},\\
\\
S^{-1}_{\omega_{3}} (p^2)
=\displaystyle{\frac{\alpha}{6\pi^2 f_{\omega_{3}}^2}\, p^4
+\frac{A_{\omega_{3}}}{f_{\omega_{3}}^2}\, p^2
-\frac{B_{\omega_{3}}}{f_{\omega_{3}}^2}}
\end{array}
\end{equation}
(the propagators of  $\omega_1$ and $\omega_2$ are 
identical owing to residual
$SU(2)_f$ symmetry (\ref{mms})).
For masses and decay constants of diquarks in the terms of
$A_{\omega_{1,2}}$, $B_{\omega_{1,2}}$, $A_{\omega_3}$, $B_{\omega_3}$
we obtain
\begin{equation}
\begin{array}{l} 
\label{mfmf-1}
m_{\omega_{1,2}}^2\!
=\displaystyle{\frac{3\pi^2 A_{\omega_{1,2}}}{\alpha}
\left(\sqrt{1+\frac{2\alpha B_{\omega_{1,2}}}{3\pi^2 A_{\omega_{1,2}}}}
-1\right)},\quad
f_{\omega_{1,2}}^2\!
=\displaystyle{A_{\omega_{1,2}}
\sqrt{1+\frac{2\alpha B_{\omega_{1,2}}}{3\pi^2 A_{\omega_{1,2}}}}},\\
\\
m_{\omega_3}^2
=\displaystyle{\frac{3\pi^2 A_{\omega_3}}{\alpha}
\left(\sqrt{1+\frac{2\alpha B_{\omega_3}}{3\pi^2 A_{\omega_3}}}
-1\right)},\quad
f_{\omega_3}^2
=\displaystyle{A_{\omega_3}
\sqrt{1+\frac{2\alpha B_{\omega_3}}{3\pi^2 A_{\omega_3}}}}.
\end{array}
\end{equation}

The constants $A_{\omega_{1,2}}$, $B_{\omega_{1,2}}$, $A_{\omega_3}$,
$B_{\omega_3}$, in their turn can be represented as a series of
quark masses. The constants $B_{\omega_{1,2}}$, $B_{\omega_3}$ contain
also the terms depending on the gluon condensate. 
Comparision of expressions for the constants
$A_{\omega_{1,2}}$, $B_{\omega_{1,2}}$, $A_{\omega_3}$, $B_{\omega_3}$
with analogous expressions for the constants
of the meson sector $A_\pi$,
$B_\pi$, $A_K$, $B_K$  (\ref{ABAB-2})
leads to the following identities:

\begin{equation}
\begin{array}{l} 
\label{ABAB-3}
\displaystyle{A_{\omega_{1,2}}=\frac{3}{2} A_K},\quad
\displaystyle{B_{\omega_{1,2}}=\frac{C_g}{18}+\frac{3}{2} B_K},\quad\\
\\
\displaystyle{A_{\omega_3}=\frac{3}{2} A_\pi},\quad
\displaystyle{B_{\omega_3}=\frac{C_g}{18}+\frac{3}{2} B_\pi},
\end{array}
\end{equation}
where $C_g$ is the gluon condensate. The constants
$A_\pi$, $B_\pi$, $A_K$, $B_K$ are expressed in the terms of dynamical
characteristics of mesons by formulae (\ref{ABAB-1}). Substituting 
(\ref{ABAB-1}) in (\ref{ABAB-3}) and using expressions 
(\ref{mfmf-1}) for 
$A_{\omega_{1,2}}$, $B_{\omega_{1,2}}$, $A_{\omega_3}$,
$B_{\omega_3}$ in (\ref{mfmf-1}) we finally have

\begin{equation}
\begin{array}{l} 
\label{mfmf-2}
m_{\omega_{1,2}}^2\!
=\displaystyle{\frac{2\pi^2}{\alpha}f_K^2
\left(\sqrt{1+\frac{\alpha C_g}{12\pi^2 f_K^4}}-1+
\frac{\alpha m_K^2}{2\pi^2f_K^2}\right)},\quad
f_{\omega_{1,2}}^2\!
=\displaystyle{\frac{2}{3} f_K^2
\sqrt{1+\frac{\alpha C_g}{12\pi^2 f_K^4}}},\\
\\
m_{\omega_3}^2
=\displaystyle{\frac{2\pi^2 }{\alpha}f_\pi^2
\left(\sqrt{1+\frac{\alpha C_g}{12\pi^2 f_\pi^4}}-1+
\frac{\alpha m^2_{\pi}}{2\pi^2f^2_{\pi}}\right)},\quad
f_{\omega_3}^2
=\displaystyle{\frac{2}{3} f_\pi^2
\sqrt{1+\frac{\alpha C_g}{12\pi^2 f_\pi^4}}}.
\end{array}
\end{equation}

The expressions (\ref{mfmf-2}) are the main result of the paper [7]. 

The masses and decay constants of scalar diquarks
are expressed in (\ref{mfmf-2}) in terms of masses and decay constants 
of pseudoscalar mesons and the gluon condensate. 
The parameter $\alpha$ displays an influence of the ``tachyonic'' term.
Its taking into account corresponds to 
$\alpha=1$ according to (\ref{L}).

Though relations (\ref{ABAB-3}) were obtained by comparision
series up to $O(m_q^2)$ it is easy to see that these
relations and consequently formulae (\ref{mfmf-2})
are valid in all orders.
Indeed, the identities (\ref{ABAB-3}) are result of
introducing scalar diquarks in the effective
low-energy theory together with pseudoscalar mesons
by means of E$\chi$-field.
The specific structure of E$\chi$-field induces a
co-ordinated influence of quark masses on dynamical parameters
of fields that is expressed in particular as identities
(\ref{ABAB-3}) and in the end relations
(\ref{mfmf-2}).

The expressions (\ref{mfmf-2}) allow to estimate
dynamical parameters of scalar diquarks. For values $f_\pi=132\, Mev$,
$m_\pi=139\, Mev$, $f_K=165\, Mev$, $m_K=495\, Mev$ and
$C_g=(365\div 405\, Mev)^4$ 
the formulae with included ``tachyonic'' term give
$m_{\omega_3}=310\div 360\, Mev$, $f_{\omega_3}=120\div 125\, Mev$,
$m_{\omega_{1,2}}=545\div 570\, Mev$, $f_{\omega_{1,2}}=140\div 145\,
Mev$. 

Using (\ref{mfmf-2}) it is easy to
relate masses and decay constants of mesons and diquarks 
Excluding the gluon condensate from 
(\ref{mfmf-2}) gives the following expressions:

\begin{equation}
\begin{array}{l} 
\label{mfmf-3}
\displaystyle{\frac{m_{\omega_{1,2}}^2 - m_K^2}{m_{\omega_3}^2 -m_\pi^2}
=\frac{ f_\pi^2+(\alpha/4\pi^2)(m_{\omega_3}^2 +m_\pi^2)}{
f_K^2+(\alpha/4\pi^2) (m_{\omega_{1,2}}^2 + m_K^2)}}\\
\\
\displaystyle{f_{\omega_3}^2 ={2\over 3} f_\pi^2
+\frac{\alpha}{3\pi^2}(m_{\omega_3}^2 -m_\pi^2),\quad
f_{\omega_{1,2}}^2 ={2\over 3} f_K^2
+\frac{\alpha}{3\pi^2}(m_{\omega_{1,2}} -m_K^2)}.
\end{array}
\end{equation}

The second and the third relations in (\ref{mfmf-3}) 
can also be considered as an alternative presentations
for decay constants of scalar diquarks.
In the limit when $\alpha=0$ and the tachionic term is absent
(\ref{mfmf-3}) is especially simple:
\begin{equation}
\begin{array}{l} 
\label{mfmf-4}
f_\pi^2 m_{\omega_3}^2-f_K^2 m_{\omega_{1,2}}^2=
f_\pi^2 m_\pi^2-f_K^2 m_K^2,\qquad
f_{\omega_{1,2}}^2=\displaystyle{\frac{2}{3}} f_K^2,\\
\\
f_{\omega_3}^2=\displaystyle{\frac{2}{3}} f_\pi^2.
\end{array}
\end{equation}

The last relation for $f_{\omega _3}$
has already been obtained [3] from a calculation of
colour degrees of freedom. In [3] the constants  $g_{\pi}$ and $g_{\omega}$
are determined as
\begin{equation}
g_\pi =<0|\bar u\gamma_5 d|\pi^+> \ \ \ \
\sqrt{\frac{2}{3}}g_\omega\delta_{ad}=
<0|\varepsilon^{abc}u^T_b C\gamma_5d_c|\omega_d>
\end{equation}
and the relation $g_\pi = g_\omega$
was obtained. We can also obtain $g_\pi$ and $g_\omega$
considering an interaction of
the chiral field with an external pseudoscalar field $P$. As a result we have

\begin{equation}
g_\pi=f_\pi\frac{m_\pi^2}{m_u+m_d}+\frac{1}{f_\pi}p^2\frac{3M}{2\pi^2}-
\frac{\alpha}{f_\pi}\frac{m^4_\pi}{4\pi^2}\frac{1}{m_u+m_d}
\end{equation}

For $g_\omega$ the expression is

\begin{equation}
\frac{f_\pi}{f_\omega}\sqrt{\frac{2}{3}}g_\pi =g_\omega 
\end{equation}
that agrees with result of [3].

In conclusion we would emphasize that formulae (\ref{mfmf-2})
for masses and decay constants of scalar diquarks, as well as
relations (\ref{mfmf-4})
are based only on the assumption that both scalar diquarks and
pseudoscalar mesons belong to the same chiral field
of the extended chiral group $E\chi$.
Our estimates for dymamical characteristics of
scalar diquarks 
give reasonable numerical values [2],[13]  
and this is an argument in favour of
describing scalar diquarks by the
effective $E\chi$ chiral theory.
Consequently, E$\chi$-action can be used
in low-energy diquark-meson physics.

The work was supported by Russian Foundation for Fundamental Studies
grant 97-01-01186.


\centerline{\bf References}

\begin{thebibliography}{13}
\bibitem{GellMann}
		  M. Gell-Mann, Phys. Lett. 8 (1964) 214;\\
		  M. Ida, R. Kobayashi, Progr. Theor. Phys. 36 (1966) 846;  \\
		  D.B. Lichtenberg, L.J. Tassie, Phys. Rev. 155 (1967) 1601.
\bibitem{}
		  S. Fredriksson, M. Jandel, Z. Phys. C 10 (1982) 41; \\
		  S. Fredriksson, M. Jandel, T.I. Larsson, Z. Phys. C 14 (1982) 35;\\
		  A.V. Hendry, I. Hinchliffe, Phys. Rev. Lett. 36 (1976) 3453; \\
		  E. Golowich, E. Haqq, Phys. Rev. D 24 (1981) 2495; \\
					 D.B. Lichtenberg, W. Namgung, E. Predazzi and J.G. Wills,
					 Phys. Rev. Lett. 48 (1982) 1653;\\
		  C.J. Burden,  R.T. Cahill,  J. Praschifka,   Aust. J. Phys. 42
		  (1989) 161;\\
		  J. Praschifka, R.T. Cahill, C.D. Roberts, Int.J.Mod.Phys. A 4,
		  No. 18 (1989) 4929;\\
					 M. Anselmino, F. Caruso, E. Leader and J. Soares, Z.Phys C 48
					 (1990) 689;\\
					 R.D. Ball, Phys.Lett.B 245 (1990) 213.
\bibitem{Stech}
					 H.G. Dosch, M. Jamin, B. Stech, Z.Phys. C 42 (1989) 167.
\bibitem{Szczekowski}
		  M. Szczekowski, Int. J. Mod. Phys. A 4, No. 16 (1989) 3985;
\bibitem{}
		  M. Anselmino, E. Predazzi, S. Ekelin, S. Fredriksson,
		  D.B. Lichtenberg, Diquarks92-a literature survey,
		  Lulea Univ. of Tech., Sweden.
\bibitem{}
		  A.A. Andrianov,  V.A. Andrianov,  V.Yu. Novozhilov,
		  Yu.V. Novozhilov, Lett. Math. Phys. 11 (1986) 217;\\
		  A.A. Andrianov, V.A. Andrianov, V.Yu. Novozhilov,
		  Yu.V. Novozhilov, Phys. Lett. B 186 (1987) 401;\\
		  Andrianov A.A., Novozhilov Yu.V., Theor. Math. Phys. 69 No.1
					 (1986) 78.
\bibitem{}       
                  A. Korotkov and A. Pronko, Yad. Phys. (to be published)
\bibitem{NPV}
					 Yu. Novozhilov, A. Pronko and D. Vassilevich, Phys.Lett. B321
					 (1994) 425.
\bibitem{}
					 R.D. Ball, Phys.Lett.B 227 (1989) 445.
\bibitem{}
					 A.A. Andrianov, Yu.V. Novozhilov, Phys.Lett.B 153 (1985) 422.
\bibitem{Ball}
				R. Ball, Phys. Rep. 182 (1989) 1;\\
					 R. Ball and G. Ripka, preprint CERN-TH.7122, to be published
					 in Proc. of Conference on Many Body Physics, Coimbra, Portugal,
		  1993.

\bibitem{}

                 D. Kahana, V. Vogl,
                 {\em Phys. Lett.\/} {\bf B 244} (1990) 10.

\bibitem{}
                 S. Ekelin, S. Fredriksson, M. Jandel, T.I. Larsson,
                 {\em Phys. Rev.\/} {\bf 28} (1983) 257.

                 S.Ekelin, S.Fredriksson, J.Hansson
                 {\em Z.Phys.\/} {\bf C 75} (1997), 107.
\end{thebibliography}
\end{document}